\documentstyle[aps,prl,tighten]{revtex}
\input{psfig.tex}
\twocolumn[\hsize\textwidth\columnwidth\hsize\csname @twocolumnfalse\endcsname]

\begin{document}
\draft 
\title{CMB polarization as a direct test of Inflation}

\author{David N. Spergel\footnote{dns@astro.princeton.edu}}
\address{Department of Astrophysical Sciences, 
Princeton University, Princeton, New Jersey~~08544}
\author{Matias Zaldarriaga\footnote{matiasz@arcturus.mit.edu}}
\address{Department of Physics, MIT, Cambridge, Massachusetts~~02139}
\maketitle

\begin{abstract}
We study the auto-correlation function of CMB polarization anisotropies and
their cross correlation with temperature fluctuations as probe of the causal
structure of the universe. Because polarization is generated at the last
scattering surface, models in which fluctuations are causally produced on
sub-horizon scales cannot generate correlations on scales larger then $\sim
2^o$. Inflationary models, on the other hand, predict a peak in the
correlation functions at these scales: its detection would be definitive
evidence in favor of a period of inflation. This signal could be detected
with the next generation of satellites.
\end{abstract}

Temperature anisotropies in the Cosmic Microwave Background (CMB) are one of
the best probes of the early universe. CMB measurements are likely
to answer one of the fundamental questions in cosmology: the origin of the fluctuations that formed the galaxies and the large-scale structure.
If the fluctuations are consistent with current
 models of structure formation, then
accurate measurements of the fluctuations could lead
to  a precise determination of a large number of cosmological parameters \cite
{jungman}.

There are two competing sets of theories for structure formation: defect
models, where a symmetry breaking phase transition generates seeds that
form sub-horizon scale density fluctuations, and inflationary models, where
a period of superluminal expansion turns quantum fluctuations into
super-horizon density perturbations. A fundamental difference between these
two mechanisms of structure formation is that only inflation alters the
causal structure of the very early universe and is able to create
correlations on super-horizon scales.

The COBE satellite observed correlations on angles much larger than that
subtended by the horizon at decoupling $(\theta _h\sim 1.1^o)$ in the CMB
temperature. 
This does not; however, imply that there were correlations on super
horizon scales at decoupling because a time dependent gravitational
potential will produce temperature fluctuations at late-times, the 
integrated Sachs Wolfe effect (ISW). For example,  cosmic string or texture
models predict that most of the fluctuations observed by COBE were
produced at $z < 10$.

Measurements of temperature fluctuations at small scales have been suggested
as a potential test of inflation: inflationary models and most 
non-inflationary
ones predict different locations and relative heights for the acoustic
peaks \cite{huwhite}. 
Unfortunately, causality alone is insufficient to distinguish inflationary
and non-inflationary temperature power spectra: causal sources that mimic
exactly the inflationary pattern of peaks can be constructed \cite{turok}.
While the predicted CMB fluctuations  of the current family of defect models
differ significantly from inflationary predictions\cite{urostur}, it is 
useful to have model independent tests of the causal structure of the early
universe.

Polarization fluctuations are produced by Thomson scatterings during the
decoupling of matter and radiation. Thus, unlike temperature fluctuations,
they are unaffected by the ISW effect. 
Measurements of the polarization fluctuations are certain to probe the
surface of last scatter. Hence, the detection of correlated polarization
fluctuations on super-horizon scales at last scattering are a definitive
signature of the existence of super-horizon scale fluctuations, one of the
distinctive predictions of inflation.\footnote{In this letter we will consider
the correlation function in real space (ie. as a function of the
separation angle)
rather than the usual power
spectrum. By doing so we can easily express the 
causality constraint,
while it would become a set of integral constraints that the power
spectrum has to satisfy in the now more usual treatment in term of $C_l$s.}

We will work in the initially unperturbed synchronous gauge, where the
metric is given by $ds^2=a^2(\tau )[-d\tau ^2+(\delta _{ij}+h_{ij})dx^idx^j]$%
. We will consider only perturbations produced by scalar modes and will
solve the Einstein equations in the presence of sources (e.g., defects)
using the stiff approximation \cite{veerasteb}. The sources are
characterized by their covariantly conserved stress energy tensor $\Theta
_{\mu \nu }$. Before recombination, matter and radiation act as a very
tightly coupled fluid, so the evolution of fluctuations can be described by 
\begin{eqnarray}
\ddot \delta _C+{\frac{\dot a}a}\dot \delta _C &=&4\pi
G(\sum_N(1+3c_N^2)\rho _N\delta _N+\Theta _{00}+\Theta )  \nonumber \\
\dot \delta _R &=&{\frac 43}\dot \delta _C-{\frac 43}\nabla \cdot {\bf {v_R}}
\nonumber \\
\dot {{\bf {v_R}}} &=&-(1-3c_S^2){\frac{\dot a}a}{\bf v_R}-{\frac 34}%
c_S^2\nabla \delta _R,  \label{eq1}
\end{eqnarray}
where $\Theta /3$ is the pressure, $\delta _R$ and ${\bf {v_R}}$ describe
the energy density and velocity of the photon-baryon fluid and $\delta _C$
is the energy density of cold dark matter.  In synchronous gauge, the cold
dark matter has zero velocity. The sum over $N$ is carried out over all
species and $c_S$ is the sound speed. Temperature and polarization
anisotropies seen on the sky today depend on $\delta _R$ and ${\bf {v_R}}$
at decoupling. 

Equations (\ref{eq1}) imply that the photon-baryon fluid propagates
information at the speed of sound and thus cannot generate correlations on
scales larger than the sound horizon. Causality on the other hand implies
that the unequal time correlators of the sources $\langle \Theta _{\mu \nu
}(r,\tau )\Theta _{\mu \nu }(0,\tau ^{\prime })\rangle $ vanishes if $r>\tau
+\tau ^{\prime }$. In the absence of initial correlations, these two
conditions together imply that $\langle X|_{\tau _{*}}(\hat {{\bf n}%
}_1)X|_{\tau _{*}}(\hat {{\bf n}}_2)\rangle =0$ if $\theta _{12}>2\theta
_h\sim 2^o$, where $X={\delta }_R,{\bf v_R},\partial _i{\bf v_R}$ and $\tau
_{*}$ is the conformal time of decoupling.

In the thin scattering surface approximation, equations (\ref{eq1}) are
solved up to recombination and then the photons free stream to the observer.
The final temperature anisotropy in direction $\hat {{\bf n}}$ on the sky is 
\begin{eqnarray}
T(\hat {{\bf n}})={\frac{\delta _R}4}|_{\tau _{*}}-\hat {{\bf n}}\cdot {\bf {%
v_R}}|_{\tau _{*}}-{\frac 12}\int_{\tau _{*}}^{\tau _0}d\tau \dot h_{ij}\hat
{{\bf n}}^i\hat {{\bf n}}^j.  \label{eqn2}
\end{eqnarray}
The first two terms are evaluated at the last scattering surface and the
third term is an integral along the line of sight, the ISW effect. In
non-inflationary models, the first two terms cannot correlate temperature
fluctuations at separations larger than $2\theta _h\sim 2^o$ but because
anisotropies can be created later through the ISW effect these models can
have temperature correlations on larger angular scales.

Polarization is produced by Thomson scattering of radiation with a non zero
quadrupole moment at the last scattering surface. The photons that scattered
off a given electron came from places were the fluid had velocity ${\bf {v_R}
}$ and thus because of the tight coupling the photon distribution function
had a dipole $T_1=\hat {{\bf n}}\cdot {\bf {v_R}}$. Furthermore gradients in
the velocity field across the mean free path of the photons ($\lambda _p$)
created a quadrupole $T_2=\lambda _pn^in^j\partial _iv_{Rj}$ in the photon
distribution ``seen'' by each electron. 
This quadrupole generates polarization through Thomson scattering. 

Linear polarization is described by a $2\times 2$ traceless tensor fully
specified by the $Q$ and $U$ Stokes parameters \cite{circular}. These
parameters depend on the direction of observation $\hat {{\bf n}}$ and on
the orientation of the coordinate system perpendicular to $\hat {{\bf n}}$, $%
(\hat {{\bf e}}_1,\hat {{\bf e}}_2)$ used to define them. Two independent
combinations with spin $\pm 2$ provide a more convenient description, $Q\pm
iU$. Under rotation of the $(\hat {{\bf e}}_1,\hat {{\bf e}}_2)$ basis by an
angle $\psi ,$ this combinations transform as $(Q^{\prime }\pm iU^{\prime
})=\exp ({\mp 2i\psi })\ (Q\pm iU)$, and can be expanded in spin $\pm 2$
harmonics, $(Q\pm iU)(\hat {{\bf n}})=\sum_{lm}a_{\pm 2,lm}\ \ _{\pm
2}Y_l^m(\hat {{\bf n}})$ \cite{spinlong}. An equivalent expansion using
tensors on a sphere can be found in \cite{kks}.
The
scattered radiation field is given by $(Q+iU)=-3/4\sigma _T\int d\Omega
^{\prime }/4\pi ({\bf m}\cdot \hat {{\bf n}}^{\prime })T_2(\hat {{\bf n}%
}^{\prime })\propto \lambda _p{\bf m}^i{\bf m}^j\partial
_iv_j|_{\tau_*}$, 
where $%
\sigma _T$ is the Thomson scattering cross section and we have written the
scattering matrix as $P({\bf m},\hat {{\bf n}}^{\prime })=-3/4\sigma _T({\bf %
m}\cdot \hat {{\bf n}}^{\prime })^2$ with ${\bf m}=\hat {{\bf e}}_1+i\hat {%
{\bf e}}_2$ . In the last step, we integrate over all directions of the
incident photons $\hat {{\bf n}}^{\prime }$.

As photons decouple from the baryons their mean free path is growing very
rapidly, so a more careful analysis is needed to get the final polarization 
\cite{zalhar}, 
\begin{eqnarray}
(Q+iU)(\hat {{\bf n}})\approx 0.17\Delta \tau _{*}{\bf m}^i{\bf m}^j\partial
_iv_j|_{\tau _{*}}  \label{eqn3}
\end{eqnarray}
where $\Delta \tau _{*}$ is the width of the last scattering surface and is
giving a measure of the distance photons can travel between their last two
scatterings. The appearance of ${\bf m}$ in equation (\ref{eqn3}) assures
that $(Q+iU)$ transforms correctly under rotations of $(\hat {{\bf e}%
}_1,\hat {{\bf e}}_2)$. Equation (\ref{eqn3}) shows that the observed
polarization only depends on the state of the fluid at the last scattering
surface. No correlations can be present in the polarization for separations
larger than $\sim 2^o$ in non-inflationary models.

Polarization can be decomposed into two distinct parts\cite{spinlong}: $E$
and $B$ with $a_{E,lm}=-(a_{2,lm}+a_{2,lm})/2$ and $%
a_{B,lm}=-(a_{2,lm}-a_{2,lm})/2i$ . Only four power spectra are needed to
describe the full radiation field, three autocorrelations $C_{Xl}={\frac
1{2l+1}}\sum_m\langle a_{X,lm}^{*}a_{X,lm}\rangle $ for $X=T,E,B$ and the
cross correlation between $E$ and $T$, $C_{Cl}={\frac 1{2l+1}}\sum_m\langle
a_{T,lm}^{*}a_{E,lm}\rangle $.

Density perturbations only contribute to $E$ \cite{spinlong,kks}. We can
illustrate this point using equation (\ref{eqn3}) and for convenience
choosing $\hat {{\bf n}}=\hat {{\bf z}}$ and ${\bf m}=\hat {{\bf x}}+i\hat {%
{\bf y}}$. In the small scale limit, we have $\nabla ^2B=(\partial
_y^2-\partial _x^2)U+2\partial _x\partial _yQ$ \cite{spinlong} (where $%
\nabla ^2=(\partial _y^2+\partial _x^2)$ is the 2-d Laplacian). This gives $%
\nabla ^2B=-\nabla ^2(\hat {{\bf z}}\cdot {\bf {\nabla }\times {v_R})}$ and
is zero because the velocity field produced by density perturbations is
irrotational.

The correlation functions of $Q$ and $U$ can be defined in a way which makes
them independent of the coordinate system \cite{kks}, given two directions
in the sky we first rotate $(\hat{{\bf e}}_1,\hat{{\bf e}}_2)$ in each
direction so that both $\hat{{\bf e}}_1$ are aligned with the great circle
connecting the two points. We then use the $Q$ and $U$ as measured in this
system to compute the correlation functions which depend only on $\theta$
the angle between the two directions. They are given by, 
\begin{eqnarray}
C^{(Q,U)}(\theta)&=&\sum_l {\frac{2l+1 }{4\pi}} [C^{(E,B)}_l\ F^1_l(\theta)
- C^{(B,E)}_l\ F^2_l(\theta) ] 
\end{eqnarray}
where $\ _{\pm 2} Y_l^2(\theta,\phi)=\sqrt{(2l+1) / 4\pi}\ [F^1_l(\theta)\pm
F^2_l(\theta)]\ \exp{(2i\phi)}$. Both correlation functions receive
contributions from the $E$ and $B$ channels.
The $E$ channel contains all the cosmological signal if there are no tensor
or vector modes.

\begin{figure}
\centerline{\psfig{figure=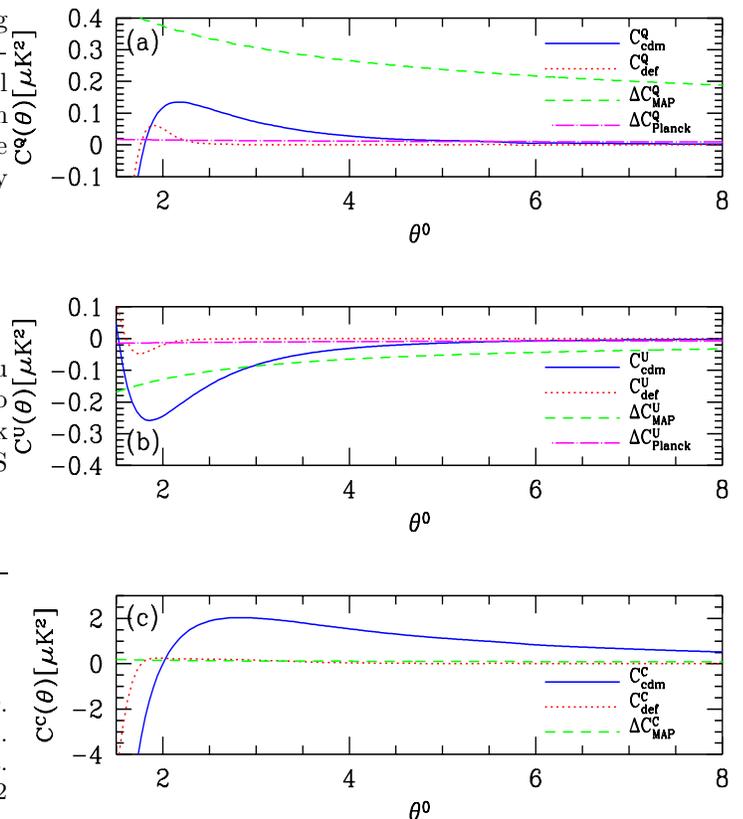,height=4.6in,width=4in}}
\caption{Correlation functions for  Q (a)  and  U (b) Stokes parameters
for sCDM and the causal seed model discussed in the text. The noise in
their determination is shown for both MAP and Planck.   Panel (b)
shows the expected noise for MAP if the CMB maps are flltered
to include only the $E$ channel signal.  Panel (c) shows
the cross correlation between temperature and polarization and the
noise for MAP, the expected variance for Planck is even smaller.
Each resolution element in the correlation function should
be considered independent.}
\label{fig1}
\end{figure}

We computed both $C^{(Q,U)}(\theta )$ for the model proposed by Turok which has a
clever choice of source stress energy tensor that is able to reproduce the
pattern of peaks of inflationary standard CDM (sCDM)\cite{turok}. 
The results are shown in figure \ref{fig1}. We see that the inflationary
model is able to produce correlations on angular scales larger than $\sim 2^o
$, while the other model cannot. On smaller angular scale than shown in
figure \ref{fig1}, the two correlation functions coincide. The difference
between the two models is a result of the causal constraints and is
insensitive to  source evolution. It is also worth pointing out that
in inflationary models the large scale polarization is suppressed relative
to the small scale signal, so we are after a small effect.   

Next, we estimate the expected uncertainties in measuring  $C^{(Q,U)}(\theta
)$. Since receiver noise is the likely to be the dominant source of
variance,  we can make a simple estimate of the total noise: it is
proportional to the number of  independent pairs of pixels, $N_p,$ at a
given separation, $\theta $. For an experiment with a full width at half
maximum of   $\theta _{fwhm},$ $N_p=1/2\times (4\pi /\theta
_{fwhm}^2)\times (2\pi \theta /\theta _{fwhm})$.  If $\sigma _{(Q,U)}^2$ is
the noise in the polarization measurement per resolution element, then the
noise in the cross correlation is given by $\Delta C^{(Q,U)}=\sqrt{2/N_p}\
\sigma _{(Q,U)}^2\approx 20\ w_P^{-1}\sqrt{0.2^o/\theta _{fwhm}}\sqrt{%
2^o/\theta }$ where $w_P^{-1}=\sigma _{(Q,U)}^2\ \Omega _{pix}/4\pi $. 

We can make a more accurate determination of the noise 
using the covariance matrix
of the different power spectra\cite{spinlong}: 
\begin{eqnarray}
{\rm Cov }(\hat{C}_{(E,B)l}^2)&=&{2\over 2l+1}(\hat{C}_{(E,B)l}+
w_P^{-1}e^{l^2 \sigma_b^2})^2
\end{eqnarray}
which give the following variances for the correlation functions,
\begin{eqnarray}
(\Delta C^{(Q,U)})^2&=&\sum_l ({2l+1 \over 4\pi})^2 
\{{\rm Cov}( C^2_{(E,B)l})\ [F^{1}_l(\theta)]^2  \nonumber \\ 
&& + {\rm Cov}( C^2_{(B,E)l})\ [F^{2}_l(\theta)]^2\}. 
\label{eqn6}
\end{eqnarray}
Figure \ref{fig1} shows the  noise in each correlation,
in the limit where
the variances are dominated by receiver noise 
$(\Delta C^{Q})^2 =
(\Delta C^{U})^2$  and
agree perfectly with our previous estimates. 
If either the cosmic 
variance is important or the power spectra of $E$ and $B$ differ,
then the approximate estimate of the previous paragraph is not
accurate and the full calculation should be used to estimate
the noise.

The noise in the correlation functions can be reduced
by focusing on the $E$-like piece of the polarization.  
The noise in the both $C^{(Q,U)}(\theta)$ 
receives contributions from the variances in both the
$E$ and $B$ spectra, but by computing both contributions separately we can
show that the  variance in $E (B)$ makes the dominant contribution to 
$\Delta C^Q (\Delta C^U)$. If we  filter the maps to pull out
only the $E$ component, then we remove not only the $B$ signal but also 
some of the noise, $\Delta C^U$ goes down 
almost by a factor of $\sim 4$.
The assumption that most of the signal is in the $E$ channel can be
checked within the data as both $E$ and $B$ contributions can be
measured separately from the maps.

For the MAP satellite,   without filtering the noise,
$\Delta C^{(Q,U)}\approx 0.36/\sqrt{\theta }\ \mu K^2$, so it will
not be sensitive enough to detect this signal, even if we combine
all of the three highest frequencies.   However,
if we filter the map to extract the E channel signal, then
the noise in the MAP 
experiment drops to 
$\Delta C^U \sim 0.1/\sqrt{\theta }\ \mu K^2$, 
and the $C^U$ signal should be detectable.
The PLANCK satellite, with its very
sensitive bolometers, should be able to achieve $\Delta C^{(Q,U)}\approx
0.003/\sqrt{\theta }\ \mu K^2$ and should easily be able to detect
both $C^U$ and $C^Q$. As cosmic variance is
not the dominant contribution to the noise,  an experiment observing a small
patch of the sky could also potentially detect this signal.


The temperature-polarization cross-correlation \cite{coulson} is another
potential test of the origin of fluctuations: 
although ISW effects produce temperature fluctuations
after decoupling,  we still  do not expect correlations
between temperature and polarization on large angular
scales for defect models. 
The  correlations between  temperature and
polarization  fluctuations directions $\hat{\bf n}_1$ and $\hat{\bf
n}_2$ are, 
\begin{eqnarray}
&& \langle Q(\hat{\bf n}_1)T(\hat{\bf n}_2)\rangle = 
\langle Q_1^* T_2^* \rangle - {1\over 2} 
\int_{\tau_*}^{\tau_0} d\tau 
\hat{ \bf n}_1^i \hat{ \bf n}_1^j \langle \dot{h}_{ij}(\tau) Q^*_1 \rangle,
\label{eqn7}
\end{eqnarray}
$T^*$ stands for the first two terms in equation (\ref{eqn2}) and 
$Q^*$ is given  by equation
(\ref{eqn3}).

In the polarization temperature cross correlation, only the term involving
the line of sight integration could produce correlations on
large angular scales.  This would require 
 correlations between the late time variations of
the metric and the velocity at last scattering. For this
to occur  in defect models, they must be moving very
fast and remain coherent as they evolve from recombination to 
very late times.  As figure 1(c) shows, even Turok's causal
seed model, which mimics inflation remarkable well in
the temperature correlation does not predict
any temperature-polarization correlation.

If 
gravity waves, rather than scalar modes,
 were the dominant source of the anisotropies, then
they could, in principle, create large angular scales cross correlations.
However, if gravity waves were significant enough to create
a large signal, then they would be directly detectable 
in the B channel. 

Figure \ref{fig1}(c) shows the calculated values of the cross correlation
together with the expected noise. The signal is well
above the noise for MAP and Planck. The detection of a large angular scale cross correlation with no 
appreciable signal in the polarization $B$ channel would put very
stringent limits on the physics of models trying to mimic inflation.


There is only one caveat to our argument: reionization. 
If the universe reionized very
early, a significant fraction of the  observed polarization will come
from the rescattering of photons at late times. 
In most scenarios, the 
fraction of rescattered photons 
is thought to be less than $\sim 20\%$ \cite{haiman}. 
Reionization has  two
effects on our argument. First, it reduces the amplitude of the 
correlation function by a factor $\exp{(-2\tau)}$,
where  $\tau$ is the optical depth to decoupling
($\tau \leq 0.2$). Second, it creates further structure in the correlation
function on large angular scales. Fortunately, the effect of reionization 
can  be separated from that of the primordial anisotropies: 
it leaves a very specific signature in the power spectrum, a peak 
at very low $l$ 
that is easily distinguish from the $l^6$ dependence
expected from causality constrains alone \cite{zalreio,Huwhite97,Battye}. Because of
the form of $F_l^1(\theta)$ and $F_l^2(\theta)$, this peak produces an
almost constant positive offset in 
$C^Q$ and $C^U$ for angles $\theta\sim 2^o$. Because the offset 
in $C^U$ is positive $(F^2_l(\theta) < 0$ for $\theta \sim 2^o$ and
$l < 70$) reionization at a relatively recent epoch  can never
create the negative peak at $\theta\sim 2^o$  
predicted by inflationary models. 

There is a precise signature in $C^{(Q,U)}(\theta)$
on $\sim 2^o$ scales that would 
allow  an unambiguous test of inflation.
The signal is small, but within 
reach for the new generation of experiments. The cross correlation 
between temperature and polarization is also expected to provide 
strong constraints that could distinguish inflation
from non-inflationary models: this signal is
much larger and will be well above the noise for MAP.
The next generation of satellites or even polarization measurements
from the ground could provide a definitive test of the inflationary
paradigm in the relatively near future.  

\acknowledgements
We thank Richard Battye, Diego Harari, 
Wayne Hu  and Uro\v s Seljak for useful discussions.
MZ is grateful to the Institute for Advanced Study where most
of this work was done.  MZ is supported by NASA NAG5-2816. 
DNS acknowledges the MAP/MIDEX project
for support.

\vfil\eject


\begin{references}
\bibitem{circular} Circular polarization cannot be
generated in the early universe through the process of Thomson
scattering, so Stokes parameter $V$ is 0.
\end{references}
\end{document}